\documentclass[preprint,aps,showpacs,floats]{revtex4}
\usepackage{graphicx}

\def\lsim{\raise0.3ex\hbox{$\;<$\kern-0.75em\raise-1.1ex
\hbox{$\sim\;$}}}
\def\gsim{\raise0.3ex\hbox{$\;>$\kern-0.75em\raise-1.1ex
\hbox{$\sim\;$}}}

\begin{document}
 
\overfullrule 0pt


\title{
\vglue -2cm{\small \hfill IFUSP-DFN/02-077
\vglue -0.2cm
\hfill IFT-P.060/2002
\vglue -0.3cm
\hfill MADPH-02-1306}\\
Measuring the Spectra of High Energy Neutrinos with a Kilometer-Scale 
Neutrino Telescope
}

\author{D. Hooper$^1$}\email{hooper@pheno.physics.wisc.edu}
\author{H. Nunokawa$^2$}\email{nunokawa@ift.unesp.br}
\author{O. L. G. Peres$^3$}\email{orlando@ifi.unicamp.br} 
\author{R. Zukanovich Funchal$^4$}\email{zukanov@if.usp.br} 

\affiliation{\\ \\
$^1$ Department of Physics, University of Wisconsin, 1150 University Avenue, 
Madison, WI 53706, USA \\  
$^2$ Instituto de F\'{\i}sica Te\'orica,Universidade Estadual Paulista,
    Rua Pamplona 145, 01405-900 S\~ao Paulo, Brazil \\
$^3$ Instituto de F\'{\i}sica Gleb Wataghin, Universidade Estadual de Campinas -- UNICAMP,13083-970 Campinas, Brazil \\ 
$^4$ Instituto de F\'{\i}sica,  Universidade de S\~ao Paulo 
C.\ P.\ 66.318, 05315-970 S\~ao Paulo, Brazil}

\begin{abstract}
 We investigate the potential of a future kilometer-scale neutrino 
 telescope such as the proposed IceCube detector in the South Pole, 
 to measure and disentangle the yet unknown components of the cosmic 
 neutrino flux, the prompt atmospheric neutrinos coming from 
 the decay of charmed particles and the extra-galactic neutrinos, 
 in the 10 TeV to 1 EeV energy range.
 Assuming a power law type spectra, 
 $d\phi_\nu/dE_\nu \sim \alpha E_\nu^\beta$, 
 we quantify the discriminating power of the IceCube detector and discuss 
 how well we can determine magnitude ($\alpha$) as well as 
 slope ($\beta$) of these two components of the high energy neutrino spectrum, 
 taking into account the background coming from 
 the conventional atmospheric neutrinos. 
 
\end{abstract}

\pacs{13.15.+g,13.85.Lg,14.60.Lm,95.55.Vj,95.85.Ry} 

\maketitle
\thispagestyle{empty}
\section{Introduction}
\label{sec:intro}
 Large volume neutrino telescopes are being constructed to detect 
 high-energy neutrinos primarily from cosmologically distant sources. 
 A major challenge for these experiments will be separating 
 the contributions  coming from the different sources in 
 the observed flux. 
 In this paper, we consider 
 three different origins for high-energy neutrinos: 
 conventional atmospheric neutrinos coming from the decay of 
 charge pions and kaons, 
 prompt atmospheric neutrinos from the decay of 
 charmed particles and  neutrinos from extra-galactic sources. 
 
 Of these sources, only the conventional atmospheric 
 neutrino flux has been observed in the energy 
 range from sub-GeV up to $\sim$ TeV range~\cite{atmnuobs}.
 Currently, the conventional atmospheric neutrino flux 
 is known to about 15-20\%~\cite{review-atm}.
 The other two fluxes, although anticipated by theoretical 
 expectations, are experimentally unknown to us, 
 and their observation will be extremely important. 

 Up to about $E_\nu \sim 100$ TeV the main source of atmospheric neutrinos 
 is the decay of pions and kaons produced in the interactions of cosmic 
 rays in the Earth atmosphere.  At higher energies, these mesons will 
 interact rather than decay, making the semileptonic decay of charmed 
 particles the dominant source of atmospheric neutrinos. This gives rise 
 to the so called prompt atmospheric neutrino flux which is, 
 unfortunately, subject to  large theoretical uncertainties.  
 The uncertainties in the calculation of 
 the prompt neutrino fluxes reflect not only our poor knowledge of the 
 atmospheric showering parameters, which for a given model can cause a 
 change of an order of magnitude in the fluxes, but are mostly 
 related to the model used to describe charm production at high energies, 
 which is responsible for a discrepancy up to two orders of magnitude in the 
 predictions~\cite{costa}. Typically, the energy dependence of prompt 
 neutrino flux is $d \phi_\nu/dE_\nu \sim E_\nu^{-3}$.

 If the prompt atmospheric neutrino flux can be determined by 
 experimental data, this can provide a unique opportunity to 
 study heavy quark 
 interactions at energies not accessible by terrestrial 
 accelerators. 
 Furthermore, the characterization of the prompt component 
 of the neutrino flux will enhance the discriminating power of 
 the other components at higher energies.
 
 High energy neutrinos are also expected to be produced in astrophysical 
 sources at cosmological distances. The most conventional source 
 candidates are compact objects such as gamma ray bursts~\cite{grb} and 
 active galactic nuclei jets, called blazars~\cite{agn}. 
 In these sources, neutrinos may be generated via pion production in the 
 collision between protons and photons in highly relativistic shocks.        
 A typical energy dependence of the extra-galactic neutrino flux in 
 these scenarios is $d \phi_\nu/dE_\nu \sim E_\nu^{-2}$. For other possibile extra-galactic neutrino spectra, see, for example, \cite{dmitry}.

 Other possible sources of extra-galactic neutrinos include neutrinos generated 
 in the annihilation of weakly interacting massive particles~\cite{wimps}, 
 the propagation of ultra-high energy protons~\cite{cosmogenic} or in a 
 variety of top-down scenarios including decaying or annihilating 
 superheavy particles with GUT-scale masses~\cite{shparticles}, decaying 
 topological defects~\cite{defects}, the so-called Z-burst mechanism~\cite{z} 
 or Hawking radiation from primordial black holes~\cite{hawking}. The neutrino fluxes from compact sources, the propagation of ultra-high 
 energy protons or top-down scenarios can be tied to the observed cosmic 
 ray flux. Since a myriad of speculations exist, resolution will 
 likely be reached only by experiment. 
 Currently, only the upper bound on such high energy extra-galactic 
 neutrino flux, $E_\nu^2d\phi_\nu/dE_\nu \lsim 10^{-5}$ 
 GeV cm$^{-2}$ s$^{-1}$ sr$^{-1}$, 
 has been obtained~\cite{amandacascade}.
 For a review of high-energy neutrino sources and detection, 
 see~\cite{review}. 
 
 Many important questions regarding the origin of cosmic rays can be 
 decided by neutrino observations. The determination of an extra-galactic 
 neutrino flux will be very important for understanding 
 the nature of the sources of the ultra-high energy cosmic rays. 

 We investigate the possibility of determining the  prompt atmospheric 
 neutrino and the extra-galactic  neutrino 
 energy spectra (slope and magnitude)  using down-going 
 showers~\cite{cascade,amandacascade} 
 induced by neutrinos in a kilometer-scale neutrino telescope 
 conceived to detect  high-energy neutrinos at high rates, 
 such as IceCube,  particularly in the region 
 $ 10 \mbox{ TeV } \lsim E_\nu \lsim 1 \mbox{ EeV}$.
 We demonstrate that since the energy spectra of these two neutrino 
 fluxes are expected to be rather different, by using shower events 
 from which one can reconstruct the initial neutrino energy with  
 some  accuracy, IceCube will be able to determine their energy 
 spectra separately even if they co-exist. 

 The organization of this paper is as follows. 
 In Sec.~\ref{sec:icecube}, we briefly describe the presumed 
 detector setup as well as the type of neutrino events we will 
 consider.
 In Sec.~\ref{sec:analysis}, we describe the analysis method  
 and in Sec.~\ref{sec:results} we present our results.  
 Finally, Sec.~\ref{sec:conclusions} is devoted to 
 discussions and conclusions. 

\section{Neutrino-Induced Shower Events in a Neutrino Telescope}
\label{sec:icecube}

 We will assume a kilometer-scale detector with excellent energy 
 and angular resolution, such as the IceCube project at the South 
 Pole~\cite{icecube}, where strings of photo-multiplier tubes are 
 distributed throughout a natural Cherenkov medium, ice.

 We are not going to be interested for our analysis in muon events since we need to be able to determine the parent neutrino energy
 with some precision; we will rather look at shower events.   
 We are interested in neutrinos in the energy 
 range from 10 TeV to 1 EeV, so we will consider all neutrinos 
 ($\nu_e$,$\nu_\mu$,$\nu_\tau$) which interact via charged or neutral 
 current interactions within or close to the detector volume and produce a shower which can be observed by the detector. 
 
 We restrict our analysis to showers induced by down-going neutrinos, 
 so we do not have to worry about energy losses and absorption 
 in the Earth and be  equally sensitive to all neutrino flavors.  
 We assume that the detector will be able to reconstruct the 
 parent neutrino energy  from the collected shower energy 
 within a factor of about 2-3, so that the data spanning  
 five decades in energy can be subdivide into the following 
 five energy bins $\Delta E_\nu= [10^{4}-10^{5},10^{5}-10^{6},
 10^{6}-10^{7},10^{7}-10^{8},10^{8}-10^{9}]$ GeV.

 The only background comes from showers induced by 
 conventional atmospheric neutrinos, which will only play a role in the 
 first two energy bins. This background can be, in principle,  
 substantially reduced if we consider only showers initiated by neutrinos 
 with zenith angle greater than 30 degrees above the horizon, since the 
 conventional atmospheric neutrino flux is peaked in this direction, 
 while prompt and extra-galactic neutrinos have approximately flat zenith 
 angle distributions.
 Another possible way to reduce the background level is to eliminate 
 shower events which are accompanied by a muon track due to charged  
 current interactions of $\nu_\mu$ with the ice. At these energies 
 the conventional atmospheric neutrino flux is mostly $\nu_\mu$
 while the prompt and extra-galactic neutrino fluxes are also expected to 
 present  a large amount of $\nu_e$  and $\nu_\tau$.
 The ratio of showers to muon tracks at a given zenith angle can also 
 be used as a way to deplete the number of background events. 
 We mention these as possible improvements to our results, but will not 
 attempt to implement them here since this type of calculations highly 
 depends on the shower angular resolution, detector acceptances and 
 efficiencies which are currently unknown.

 We estimate the number of neutrino-induced showers in the 
 $i$-th bin, $N_{i}$,  in a kilometer-scale detector 
 simply by 

\begin{equation}
N_i = \displaystyle A \int_{\Delta E_\nu^i} \frac{d\phi_\nu}{dE_\nu} \, \sigma_\nu(E_\nu) dE_\nu \, d\Omega,
\label{nevt}  
\end{equation}
 where  $A=N_A \times T \times V \times \rho$, $N_A$ being the Avogadro's 
 number,  
 $T$ the exposure time of observation, 
 $V$ the detector effective  volume (assumed to be 1 km$^3$)
 $\rho \sim 1$ g/cm$^3$, the ice density.
 The  neutrino interaction cross section~\cite{gandhi}, $\sigma_\nu$, includes 
 charged and neutral current contributions and the neutrino flux. 
 $d\phi_\nu/dE_\nu$ will vary according to our theoretical assumptions 
 for the flux energy dependence. Integration over the upper hemisphere 
 as well as average in each energy bin is implied.
 
 We parametrize the extra-galactic or the prompt neutrino flux 
 spectrum  by two parameters $(\alpha, \beta)$ as,  
\begin{equation}
\frac{d\phi_\nu}{dE_\nu} \equiv \alpha \left(
\frac{E_\nu}{E_0}
\right)^\beta,
\label{param}
\end{equation}
 where we fixed $E_0= 1$  GeV
 and $\alpha$ is defined to be given in units of 
GeV$^{-1}$ cm$^{-2}$ s$^{-1}$ sr$^{-1}$ throughout this paper. 
Roughly speaking, it is expected that, 
$\beta \sim -2$ and $-3$ for extra-galactic and 
prompt atmospheric neutrino flux, respectively. 
In this work, we assume that we do not know, {\it a priori}, 
the spectrum index but try to determine it experimentally.  

 For the calculations of the number of conventional atmospheric 
 neutrino shower events in the $i$-th bin, $N_i^{\text{atm}}$,
 we substitute  $d\phi_\nu/dE_\nu$ in Eq.~(\ref{nevt}) by the  
 Bartol flux~\cite{nuflux} which will be considered to be the reference 
 conventional atmospheric neutrino flux in this paper. 

\section{Analysis Method}
\label{sec:analysis}

 In order to quantify the discriminating power of IceCube type detectors 
 to different flux models, we use the $\chi^2$ function which is defined 
 as 

\begin{eqnarray}
\chi^2  & \equiv  &  \displaystyle
\sum_{i=1}^{2} \frac{\left(N_i^{\text{th}}-N_i^{\text{obs}}\right)^2}
{N_i^{\text{obs}}+N_i^{\text{atm}}+ (\sigma_{\text{atm}}  
N_i^{\text{atm}})^2}  \nonumber  \\ 
  & + &  2 \times  \sum_{i=3}^{5}
\left\{N_i^{\text{th}}-N_i^{\text{obs}}+N_i^{\text{obs}}
\ln[N_i^{\text{obs}}/N_i^{\text{th}}]\right\}, 
\end{eqnarray}
 where $N_i^{\text{th}}=N_i(\alpha_{\text{EG}},\beta_{\text{EG}}),
 N_i(\alpha_{\text{prompt}},\beta_{\text{prompt}})$ or the sum, 
 $N_i^{\text{obs}}=N_i(\alpha^0_{\text{EG}},\beta^0_{\text{EG}}),
 N_i(\alpha^0_{\text{prompt}},\beta^0_{\text{prompt}})$ or the sum,  
 and the energy bins, with energy varied from 10 TeV to 1 EeV, are 
 as defined in Sec.~\ref{sec:icecube}.  Note that the $\chi^2$ 
 will be either a function of two or four variables.

 The conventional atmospheric neutrino flux has to be considered as 
 a background to the observation of any other component up 
 to $E_\nu \sim 1$ PeV. We assume the conventional atmospheric 
 flux prediction can be subtracted from the data and include the 
 statistical ($\sqrt{N_i(\alpha^0,\beta^0)}$ and 
 $\sqrt{N_i^{\text{atm}}}$)
 as well as the systematical ($\sigma_{\text{atm}} N_i^{\text{atm}}$)
 uncertainties coming from this data in the $\chi^2$ for the first two bins.
 We note that $\sigma_{\text{atm}}$ indicates the theoretical 
 uncertainty in the absolute normalization of the conventional 
 atmospheric neutrino flux which can be significantly reduced 
 by future measurements.

 The analysis strategy we propose is the following. 
 In the future, when data exists, the spectrum should be first fitted with a single power law type spectrum. 
 If it can be well fitted by such a power law with 
 $\beta \sim -2$ $(-3)$ we will be able to 
 conclude that the data is most likely dominated by extra-galactic (prompt atmospheric) neutrinos. If they can not be well fitted by a single power law spectrum, the next step should be to fit them with two components with 
 different power laws. 
 
 Since we do not yet have sufficient data, 
 we will simulate an experimental data set which either have 
 pure or dominant extra-galactic, pure or dominant 
 prompt or a combination of extra-galactic and prompt neutrino components. 
 Then we will perform a $\chi^2$ fit to see if we can correctly 
 reproduce the input values, without any assumption about 
 these parameters. 

 For a given input, 
 we first try to fit the simulated data with a single component, {\em i.e.},  
 by minimizing  $\chi^2 (\alpha,\beta)$. If this fit is not very good,
 $\chi^2_{\text{min}} \gsim 11.8$, then we try to perform a two 
 component fit, {\em i.e.}, by minimizing  $\chi^2(\alpha_{\text{EG}},
 \beta_{\text{EG}},\alpha_{\text{promt}},\beta_{\text{prompt}})$. 
 After minimizing the $\chi^2$ function, we calculate the 
 allowed region in the $\alpha \times \beta$ plane by imposing  
 $\chi^2(\alpha,\beta)=\chi^2_{\text{min}}+ 11.8$, which 
 corresponds to a 3 $\sigma$ level estimation.

\section{A Three Prong View of the Problem}
\label{sec:results}

\begin{figure}
\centering\leavevmode
\hglue -1.5cm
\includegraphics{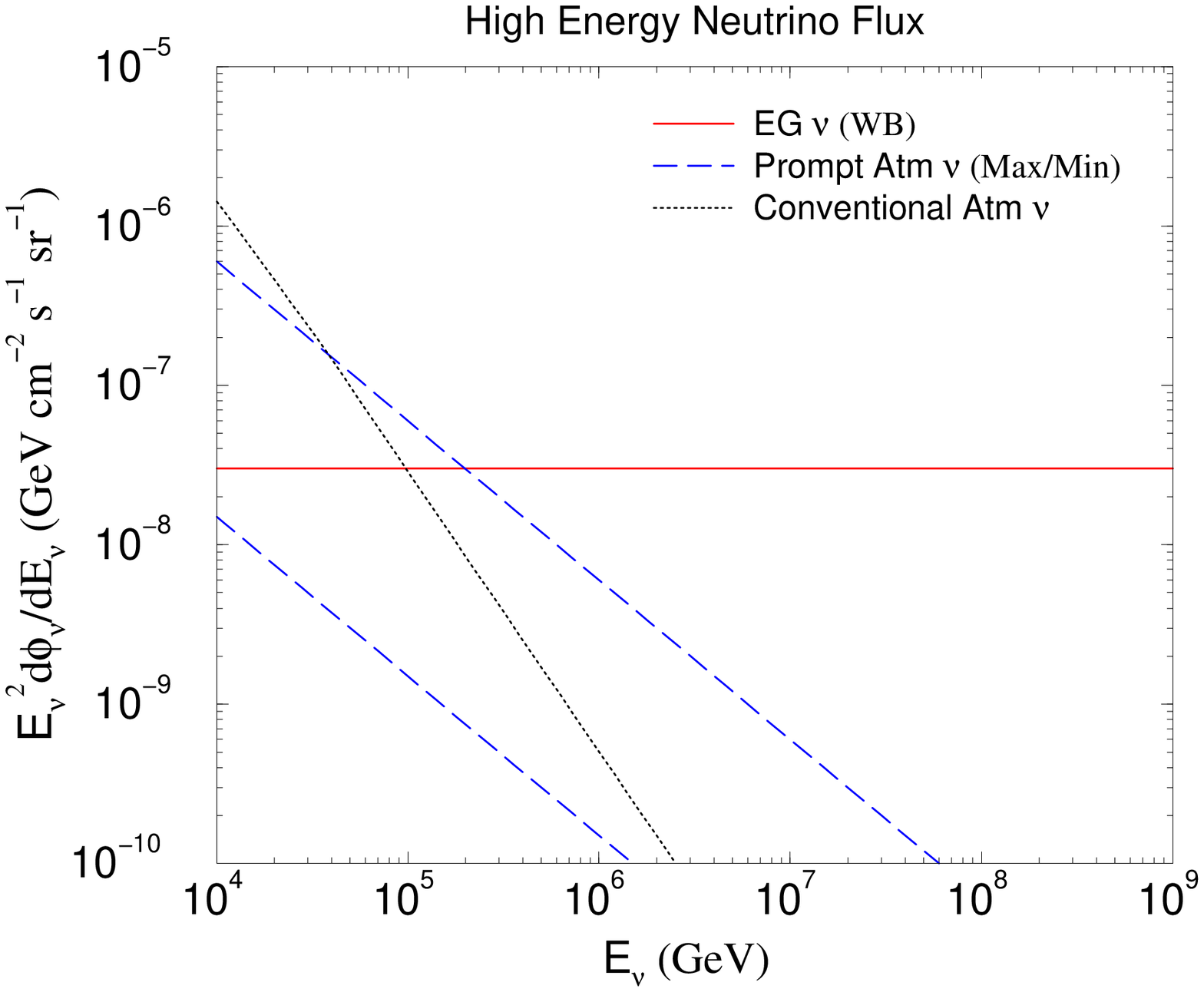}
\vglue -0.1cm
\caption{
 The three kind of high energy neutrino fluxes that are 
 considered in this paper: 
 conventional atmospheric neutrinos (dotted line), 
 the Waxman-Bahcall (WB) flux for extra-galactic 
 neutrinos (solid line) and 
 prompt atmospheric neutrinos (region circumscribed 
 by the dashed lines).  
}
\label{fig1}
\end{figure}

 We first show in Fig.~\ref{fig1} the theoretical expectations for 
 the three contributions to the neutrino flux we will be 
 considering in this work. The conventional atmospheric neutrino 
 flux has currently a theoretical uncertainty of about 15\%. 
 The prompt neutrino contribution is only known within 2 orders of 
 magnitude, its minimum and maximum values are shown in the plot by 
 the dashed lines, which rougly correspond to the range
 discussed in ~\cite{costa}.
 The Waxman-Bahcall (WB) flux \cite{WB}, which is 
 shown in the plot, will be considered to be our reference 
 extra-galactic neutrino flux.

 We show in Fig.~\ref{fig2}, the expected number of shower events
 for the neutrino fluxes presented in Fig.~\ref{fig1}. 
 As expected from Fig.~\ref{fig1}, the contribution from 
 conventional atmospheric neutrinos dominates in the 1st energy 
 bin and then it drops very quickly as energy increases.  
 Because of the weak slope ($\beta = -2$), the contribution from 
 extra-galactic neutrinos drops slowly as the energy increases and 
 the flux from prompt neutrinos drops faster 
 than the flux from extra-galactic neutrinos but slower than 
 that from the conventional atmospheric neutrinos. 
 From this plot, we can anticipate that the energy spectra ($\beta$) 
 of extra-galactic and prompt neutrinos can be determined experimentally 
 with certain accuracy. 
 Below, we will quantify the precision of the 
 determination of the flux parameters for various cases.

\subsection{Assuming a Dominant Extra-Galactic Component in the Data}
\label{sec:extrag}

 Let us first discuss the case where extra-galactic neutrino contribution 
 is much larger than the prompt neutrino flux. 
 In Fig.~\ref{fig3} we show how well the extra-galactic 
 neutrino flux component can be 
 determined by IceCube, after 1 and 10 years of data taking, for 
 two other values of $\alpha^0_{\text{EG}}$ besides the reference 
 WB ($\alpha^0_{\text{EG}}=3\times10^{-8}$ GeV$^{-1}$ cm$^{-2}$ s$^{-1}$ sr$^{-1}$) one.

\begin{figure}
\vglue -2.5cm
\hglue -1.5cm
\centering\leavevmode
\includegraphics{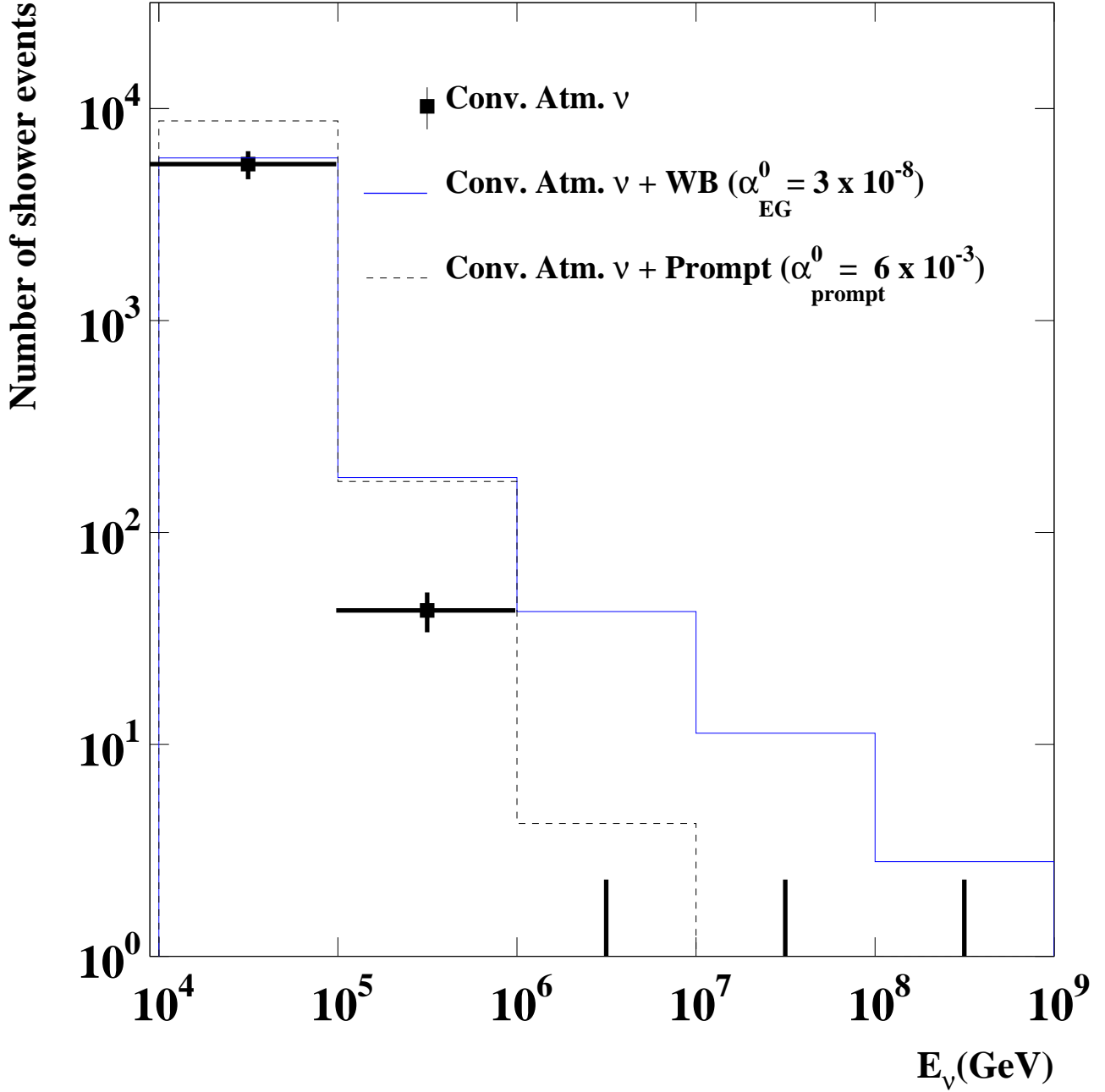}
\vglue -5.5cm
\caption{
 Number of down going shower events expected after 10 years 
 of data taking in IceCube. 
 The number of conventional atmospheric neutrino showers 
 are shown with an error bar which includes the statistical as well 
 as the systematic error with $\sigma_{\text{atm}}=15$\%.
Two values of assumed input $\alpha^0$'s are indicated in 
the plot in units of GeV$^{-1}$ cm$^{-2}$ s$^{-1}$ sr$^{-1}$. 
}
\label{fig2}
\end{figure}

\begin{figure}
\centering\leavevmode
\vglue -0.7cm
\hglue -1.5cm
\includegraphics{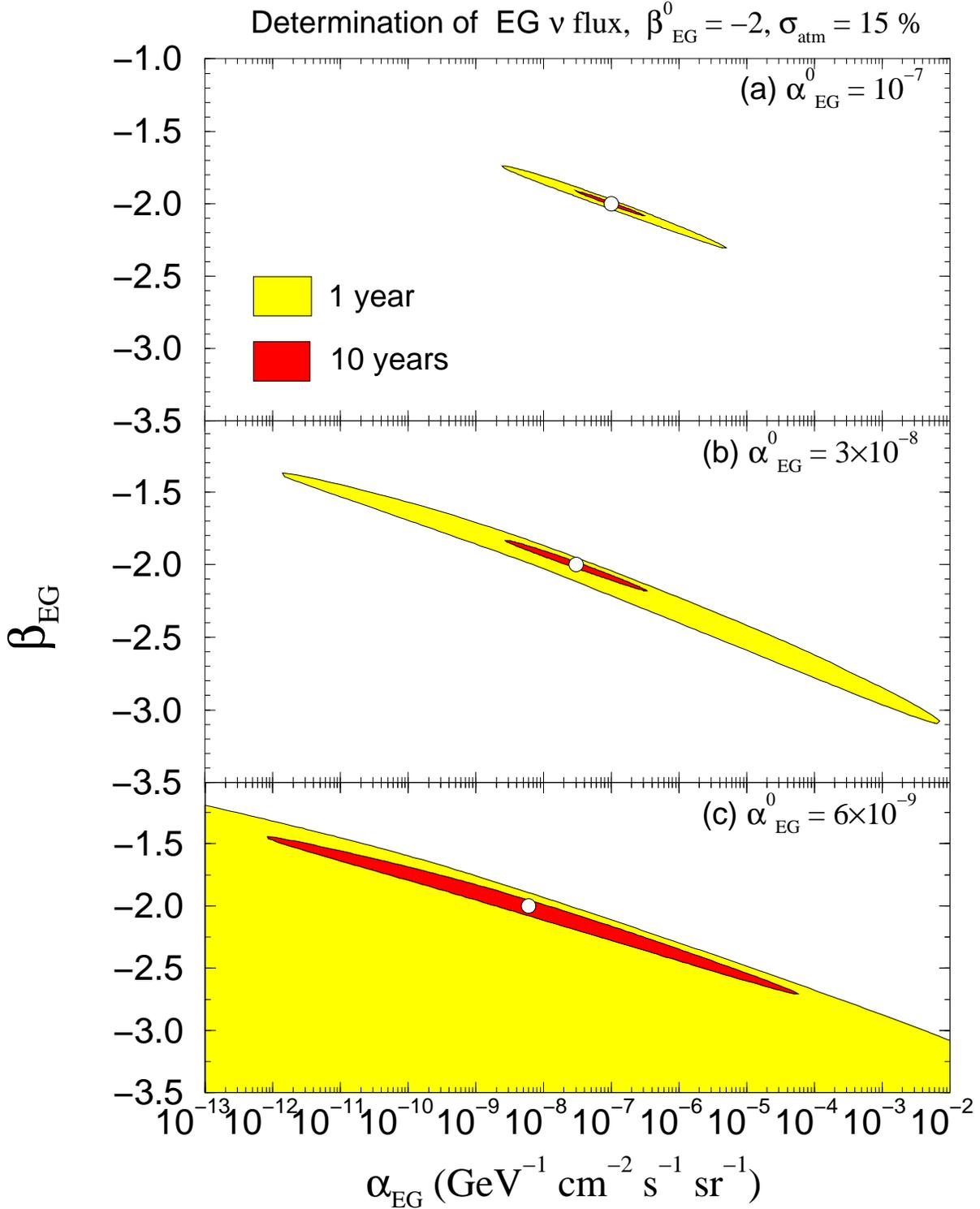}
\vglue 1.1cm
\caption{
 Determination of the parameters 
 $(\alpha_{\text{EG}},\beta_{\text{EG}})$, after 1 and 10 years 
 of IceCube observations, in the case the data is consistent with 
 a dominant extra-galactic neutrino flux. 
 We have assumed as an input $\beta^0_{\text{EG}}=-2$ 
 and $\alpha^0_{\text{EG}}$(GeV$^{-1}$ cm$^{-2}$ s$^{-1}$ sr$^{-1})=$  
 $10^{-7}$ (a), $3 \times 10^{-8}$ (b) and $6 \times 10^{-9}$ (c).  
 The contours represent the determination at 3 $\sigma$ level.
 Contributions from the conventional atmospheric neutrinos are included 
 in the $\chi^2$ with $\sigma_{\text{atm}}=15$\%.
}
\label{fig3}
\end{figure}

 We have found that if the major component of the data are events 
 induced by neutrinos coming from astrophysical sources, due to the 
 difference in the slope of the conventional atmospheric neutrino flux 
 and the extra-galactic flux, the first energy bin is only important  
 for the determination of the flux parameters in the first year of data 
 taking. After 10 years this contribution is completely irrelevant, 
 which means that the events in the first bin can be completely ignored 
 (a fit with four bins would be just as good), see Fig.~\ref{fig2} 
 where we plot the number of shower events per energy bin. 
 This also imply that our results are independent of the magnitude of the  
 theoretical systematic error assumed for the conventional atmospheric 
 neutrino calculation. 

 On the other hand, for the determination of the maximal sensitivity 
 of IceCube, the background from conventional atmospheric events in 
 the second bin is important and (see Fig.~\ref{fig2}), in this case,
 there is some dependence on the value assumed for the systematic 
 error.

 We have calculated that after 10 years of observations, IceCube 
 will be able to determine $\alpha_{\text{EG}}$ within an order of magnitude 
 and $\beta_{\text{EG}}$ to $\approx$ 10\%, assuming as input a dominant WB flux. 
 We have also estimated that the maximal sensitivity of 
 IceCube after 10 years of data taking to be
 $\alpha^0_{\text{EG}} \approx 6 \times 10^{-9}$ 
 GeV$^{-1}$ cm$^{-2}$ s$^{-1}$ sr$^{-1}$.

\subsection{Assuming a Dominant Prompt Component in the Data}
\label{sec:prompt}

 Next let us consider the case where the prompt neutrino  
 component is dominant. 
 As can be seen in Fig.~\ref{fig2} the number of prompt neutrino 
 shower events drops drastically after the second energy bin. 
 This makes the determination of this flux, even if dominant over 
 the extra-galactic flux, very sensitive to the theoretical uncertainty 
 in the conventional atmospheric neutrino flux determination.

 In fact, the flux determination will basically rely on the number 
 of shower events in the first two energy bins.
 Since the 1st bin suffers from the background from the conventional
 atmospheric neutrino flux, we can only explore a relatively narrow 
 range in $\alpha^0_{\text{prompt}}$ and, as a general rule, 
 the parameters 
 $\alpha_{\text{prompt}}$ and $\beta_{\text{prompt}}$  can be 
 at most determined within 2 orders of magnitude and about 20\%,  
 respectively, with the present value of $\sigma_{\text{atm}}=$ 15\%. 

\begin{figure}
\centering\leavevmode
\vglue -0.7cm
\hglue -1.5cm
\includegraphics{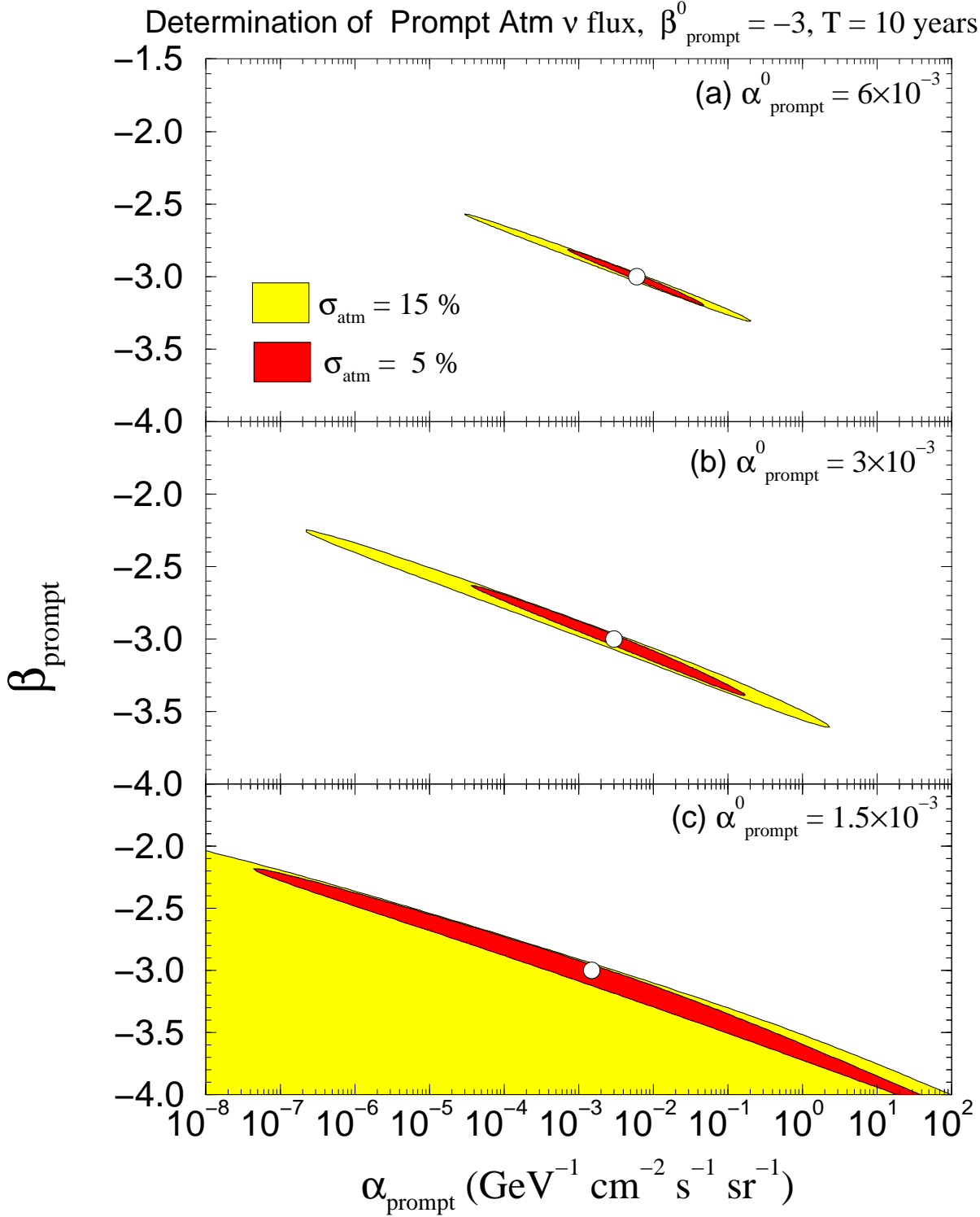}
\vglue 1.1cm
\caption{
 Determination of the parameters $(\alpha_{\text{prompt}}, 
 \beta_{\text{prompt}})$, 
 after 10 years of IceCube observations, in the case the data 
 is consistent with a dominant prompt neutrino flux.
 We have assumed as an input $\beta^0_{\text{prompt}}=-3$ and 
 $\alpha^0_{\text{prompt}}$(GeV$^{-1}$ cm$^{-2}$ s$^{-1}$ sr$^{-1})=$  
 $6 \times 10^{-3}$ (a), 
 $3 \times 10^{-3}$ (b) and $1.5 \times 10^{-3}$ (c).  
 The contours are determined at 3 $\sigma$ level.
 Contributions from the  conventional atmospheric neutrinos are 
 included in the $\chi^2$ with $\sigma_{\text{atm}}=15$ and 5\%.
}
\label{fig4}
\end{figure}

  To illustrate the effect of the systematical error $\sigma_{\text{atm}}$, 
  we show in Fig.~\ref{fig4} how well the parameters of the prompt 
  neutrino flux component can be determined by IceCube, 
  after 10 years of data taking, 
  for $\sigma_{\text{atm}}=$ 15 and 5\% and for three possible values of 
  $\alpha^0_{\text{prompt}}$ (GeV$^{-1}$ cm$^{-2}$ s$^{-1}$ sr$^{-1}$): 
  (a) $6\times 10^{-3}$, which corresponds to the maximum allowed 
  value by the theoretical calculations (see Fig.~\ref{fig1}); 
  (b)$3\times 10^{-3}$ and  
  (c) $1.5\times 10^{-3}$, where we clearly reach the maximal 
  sensitivity of IceCube. 

\subsection{Disentangling Extra-galactic and Prompt Components}
\label{sec:des}
 
 Finally, let us consider the case where both 
 extra-galactic and prompt components give significant contributions. 
 In order to determine whether it is possible to disentangle these  
 two yet not measured components of the cosmic neutrino flux, if 
 they are equally present in the data, we have investigated if it 
 would be possible to fit the measured flux with a single power law 
 assuming the data would be consistent with various values of 
 $\alpha^0_{\text{EG}}$ and $\alpha^0_{\text{prompt}}$. 
 We were able to compute the region in the 
 ($\alpha^0_{\text{EG}},\alpha^0_{\text{prompt}}$) plane which cannot 
 be explained by a single power law for different assumptions on 
 $\sigma_{\text{atm}}$. This was done by projecting in this plane the 
 3 $\sigma$ level region which corresponds to 
 $\chi^2_{\text{min}}(\alpha^0_{\text{EG}}, \alpha^0_{\text{prompt}})+ 11.8$. 
 This is shown in Fig.~\ref{fig5}. 

 We see that for $\sigma_{\text{atm}}=$ 15\%, 
 $\alpha^0_{\text{EG}}=3\times 10^{-8}$ GeV$^{-1}$ cm$^{-2}$ s$^{-1}$ sr$^{-1}$
 (our reference value) and $\alpha^0_{\text{prompt}}= 6 \times 10^{-3}$ 
 GeV$^{-1}$ cm$^{-2}$ s$^{-1}$ sr$^{-1}$ (the maximal allowed value for 
 the prompt neutrino flux) is a critical point, just on the boundary. 

 If the uncertainty in the overall normalization on the conventional 
 atmospheric neutrino flux do not decrease by future measurements, 
 this means that it will be very difficult to say 
 anything definite about the prompt neutrino flux, assuming extra-galactic 
 neutrinos  also contribute to the data. In this case the two components 
 will be indistinguishable and the extra-galactic neutrino 
 flux will dominate the fit.
 For more optimistic values of $\sigma_{\text{atm}}$, the situation 
 improves, so if  $\sigma_{\text{atm}}=$ 5\% can be achieved 
 the prompt neutrino flux can be separated from the WB neutrino 
 flux for $\alpha^0_{\text{prompt}} \gsim 2 \times 10^{-3}$  
 GeV$^{-1}$ cm$^{-2}$ s$^{-1}$ sr$^{-1}$.

 One would expect that an increase of  $\alpha^0_{\text{EG}}$ with 
 a corresponding decrease of $\alpha^0_{\text{prompt}}$ or 
 a decrease of $\alpha^0_{\text{EG}}$ with a corresponding increase 
 of $\alpha^0_{\text{prompt}}$ would help to separate the fluxes.  
 This is in fact observed in Fig.~\ref{fig5}. Nevertheless, 
 as the extra-galactic neutrino flux increases, lower values of 
 the prompt neutrino flux can be 
 distinguished up to a minimum, where the prompt neutrino flux 
 and the conventional atmospheric neutrino flux become virtually equal 
 and indistinguishable as background. There is also a minimum value for 
 the extra-galactic neutrino flux, below which the statistics 
 are too low to be disentangled. 
  
 To illustrate the impact of $\sigma_{\text{atm}}$, we show in 
 Fig.~\ref{fig6}  how well the prompt and extra-galactic components 
 can be determined by IceCube, after 10 years of data taking, 
 for $\sigma_{\text{atm}}= $ 10 and 5\%. In both cases the 
 two components can be well separated, as expected from Fig.~\ref{fig5}, 
 but $\alpha_{\text{EG}}$, $\alpha_{\text{prompt}}$  will be determined 
 within 3-4 orders of magnitude, $\beta_{\text{EG}}$ within about 20-30\%  
 and $\beta_{\text{prompt}}$ within about 20-40\%.   

 For the critical point  $\alpha^0_{\text{EG}}=3\times 10^{-8}$ 
 GeV$^{-1}$ cm$^{-2}$ s$^{-1}$ sr$^{-1}$ and 
 $\alpha^0_{\text{prompt}}= 6 \times 10^{-3}$ 
 GeV$^{-1}$ cm$^{-2}$ s$^{-1}$ sr$^{-1}$ of Fig.~\ref{fig5} we have 
 investigated the correlation between the determination of 
 $\beta_{\text{EG}}$ and $\beta_{\text{prompt}}$, for 
 $\sigma_{\text{atm}}=$ 15, 10 and 5\%. In Fig.~\ref{fig7} 
 we show the corresponding allowed regions projected in this plane. 
 From this figure it is clear why at  $\sigma_{\text{atm}}=$ 15\% 
 the single power law fit is still marginally acceptable. In this case the 
 region allowed at 3 $\sigma$ touches 
 the $\beta_{\text{EG}}=\beta_{\text{prompt}}$ line, so 
 this possibility cannot be completely discarded.
 Any improvement on $\sigma_{\text{atm}}$ will place this point 
 out of the allowed region, making the single power law fit 
 unsuitable to explain the data.

\section{Discussions and Conclusion}
\label{sec:conclusions}

 We have investigated the possibility of future neutrino telescopes to separate
 the various contributions to the observed neutrino flux. We have 
 considered that high-energy neutrinos from three different origins 
 can contribute to the measured flux: conventional atmospheric 
 neutrinos, prompt atmospheric neutrinos from the decay of charmed 
 particles and  neutrinos from extra-galactic sources. 

 We have restricted our analysis to showers induced by down-going neutrinos, 
 not to have to worry about energy losses in the Earth and be 
 equally sensitive to all neutrino flavors.  
 We have also assumed the neutrino telescope will be able to 
 reconstruct the  parent neutrino energy  from the collected shower energy 
 within a factor of about 2-3.   

 Assuming the prompt atmospheric and extra-galactic neutrino fluxes 
 can be described by a power law and 
 parametrized by  two parameters $\alpha$ (the magnitude) and $\beta$ 
 (the slope), and considering that the conventional atmospheric 
 neutrino flux 
 is currently known with a theoretical uncertainty $\sigma_{\text{atm}}=$
 15\%, our conclusion are the following.

 If extra-galactic neutrinos constitute the dominant component of the measured 
 flux, after 10 years of observations, a detector such as IceCube 
 will be able to determine $\alpha_{\text{EG}}$ within an order of magnitude 
 and $\beta_{\text{EG}}$ to $\approx$ 10\%, assuming as input a dominant WB 
 flux. This is independent of the conventional atmospheric neutrino 
 contamination.
 We have also estimated that the maximal sensitivity of 
 IceCube after 10 years of data taking will be 
 $\alpha^0_{\text{EG}} \approx 6 \times 10^{-9}$ GeV$^{-1}$ 
 cm$^{-2}$ s$^{-1}$ sr$^{-1}$.
 
  If prompt neutrinos constitute the dominant component of the measured 
  flux, after 10 years, IceCube can determine $\alpha_{\text{prompt}}$ 
  and $\beta_{\text{prompt}}$ at most within 2 orders of 
  magnitude and about 20\%,  respectively, with the present value 
  of $\sigma_{\text{atm}}=$ 15\%. This can nevertheless be improved 
  if this uncertainty can be substantially reduced. We also have 
  estimated that in this case, the maximal sensitivity of IceCube will be  
  achieved for $\alpha^0_{\text{prompt}}=1.5\times 10^{-3}$ 
  GeV$^{-1}$ cm$^{-2}$ s$^{-1}$ sr$^{-1}$.

\begin{figure}
\centering\leavevmode
\hglue -1.5cm
\includegraphics{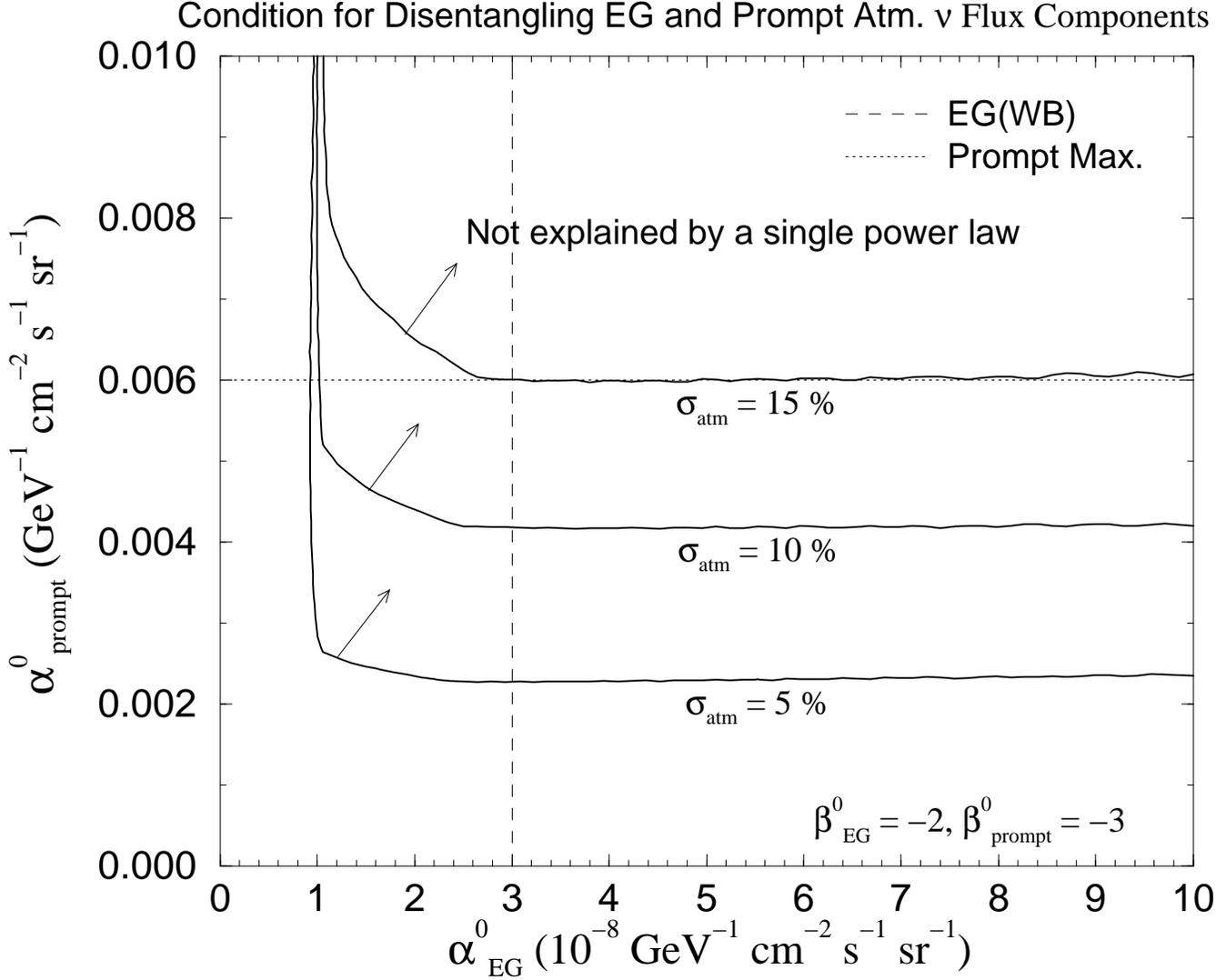}
\vglue 0.1cm
\caption{
 Region in the $(\alpha^0_{\text{EG}},\alpha^0_{\text{prompt}})$
 plane where the simulated data cannot be explained by a single power 
 law for $\sigma_{\text{atm}}=$ 15, 10 and 5\%, 
 indicated by allows. 
 The curves have been computed for $\chi^2=\chi^2_{\text{min}}+ 11.8$ 
 (3 $\sigma$ level).
 The vertical and horizontal dashed lines cross at the critical 
 point where $\alpha^0_{\text{EG}}$ takes the value for the WB flux 
 and $\alpha^0_{\text{prompt}}$ takes the maximal allowed value for 
 the prompt neutrino flux contribution in Fig.~\ref{fig1}. 
}
\label{fig5}
\end{figure}

\begin{figure}
\centering\leavevmode
\hglue -1.0cm
\includegraphics{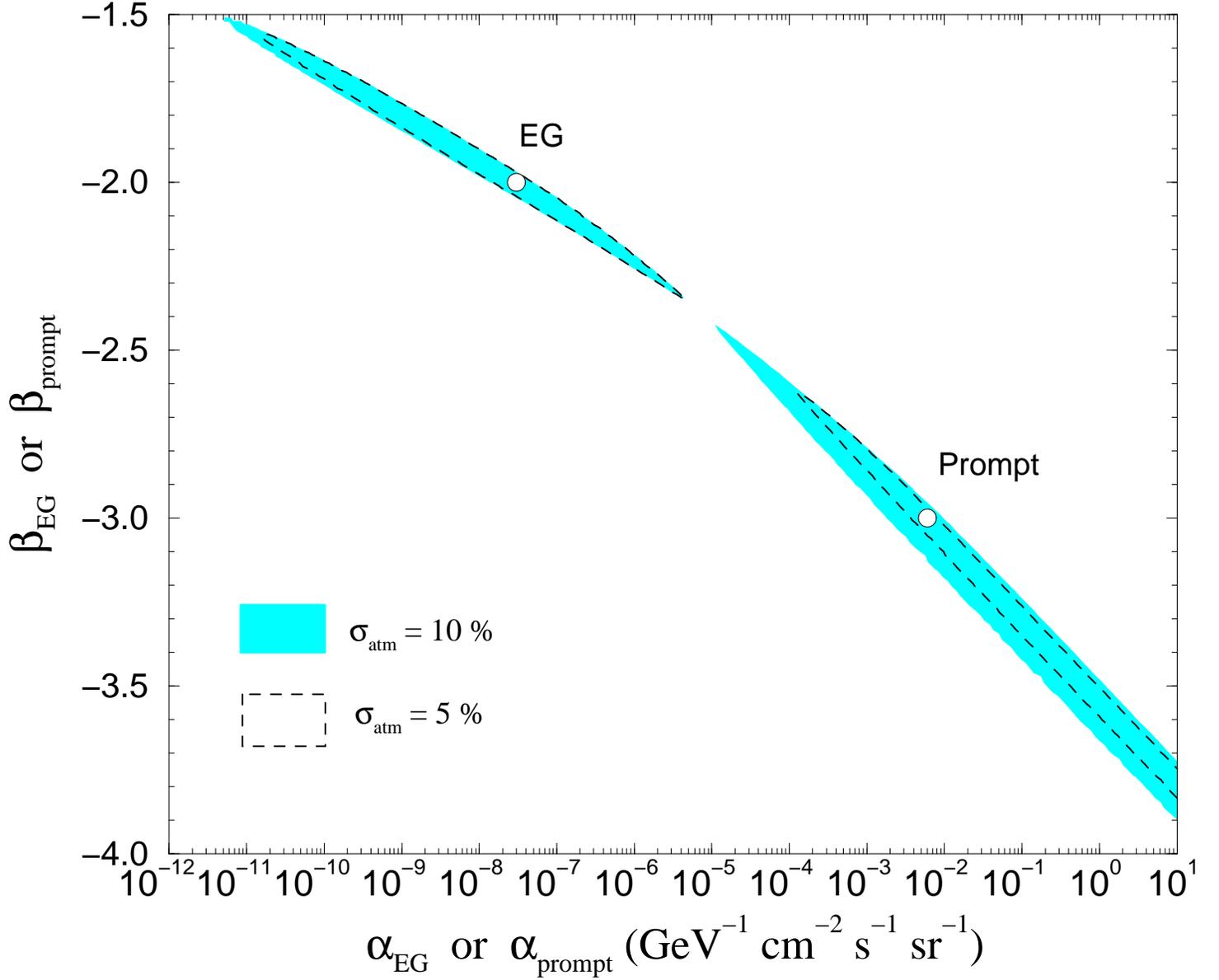}
\vglue 0.1cm
\caption{
 Same as Fig.~\ref{fig4}, but in the case that the data have 
 contributions from both prompt and extra-galactic neutrino fluxes.
 We have assumed as input $\beta^0_{\text{prompt}}=-3$, 
 $\alpha^0_{\text{prompt}}=6 \times 10^{-3}$ GeV$^{-1}$  
 cm$^{-2}$ s$^{-1}$ sr$^{-1}$ and $\beta^0_{\text{EG}}=-2$,  
 $\alpha^0_{\text{EG}}=3 \times 10^{-8}$ GeV$^{-1}$ 
 cm$^{-2}$ s$^{-1}$ sr$^{-1}$, $\beta^0_{\text{EG}}=-2$.  
 Contributions from the conventional atmospheric neutrinos are included 
 in the $\chi^2$ with $\sigma_{\text{atm}}=10$\% (shaded area) 
and 5\% (area delimited by the dashed curves). 
}
\label{fig6}
\end{figure}

\begin{figure}
\centering\leavevmode
\hglue -1.5cm
\includegraphics{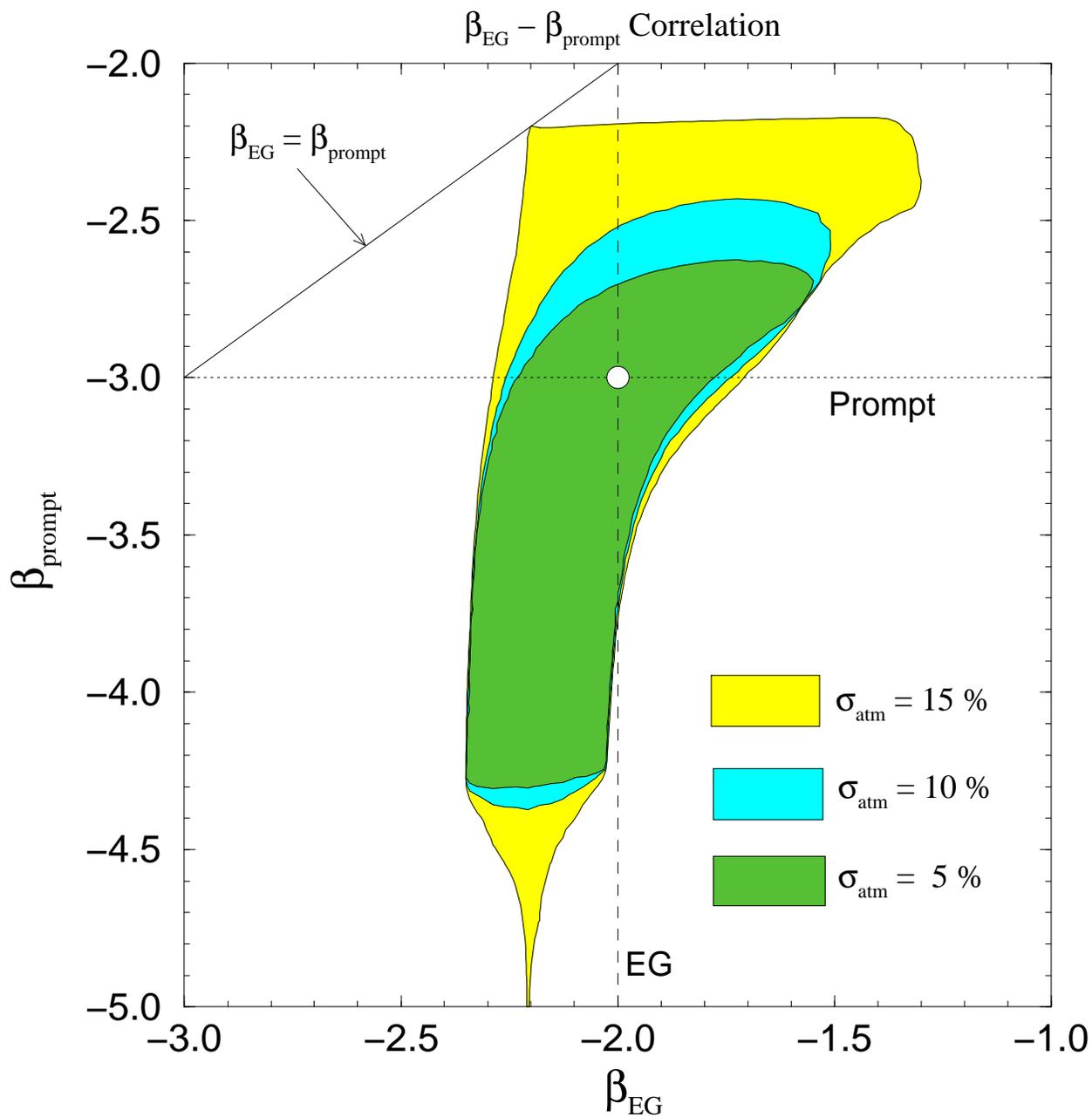}
\vglue 0.2cm
\caption{
 Allowed region in the ($\beta_{\text{EG}}$, $\beta_{\text{prompt}}$) 
 plane for the same input values of Fig.~\ref{fig6}. 
 The contours are also determined at 3 $\sigma$ level.
 Contributions from the  conventional atmospheric neutrinos are 
 included in the $\chi^2$ with $\sigma_{\text{atm}}=$ 15, 10 and 5\%.
}
\label{fig7}
\end{figure}

  We have also determined in which cases a complete separation 
  of the two components can be performed if both extra-galactic and prompt neutrinos 
  contribute to the observed flux, Fig.~\ref{fig5} summarizes our 
  conclusions on this.
  The main point here is that to clearly separate the prompt  
  component from the extra-galactic component $\sigma_{\text{atm}}$ must be about 
  10\% or less. If $\sigma_{\text{atm}}$ is much larger, 
  a single power law will fit the data with an acceptable
  value of $\chi_{\text{min}}^2$.   

  Finally, let us mention that there is an additional signature 
  that can be used to distinguish extra-galactic neutrinos from 
  the prompt atmospheric ones. 
  As first indicated by atmospheric neutrino data 
  and lately confirmed by the K2K experiment~\cite{K2K}, $\nu_\mu$ oscillate 
  to $\nu_\tau$ implying that one third of the total original extra-galactic 
  $\nu$  flux will arrive at the Earth as $\nu_\tau$. 
  On the other hand, prompt neutrinos are expected to have much 
  lower $\nu_\tau$ than $\nu_e$ or $\nu_\mu$ content~\cite{costa}.
  For $E_\nu \gsim 1$ PeV a $\nu_\tau$ event can be clearly  
  recognized through the observation a $\tau$, produced by a
  $\nu_\tau$ charge current interaction, which will 
  decay in the detector. This gives rise to the so-called 
  double-bang (when the $\tau$ is produced and decays within 
  the detector volume) and lolly pop (when the $\tau$ is produced
  outside the detector but decays inside it) events~\cite{review,pakvasa}. 
  We estimate that after 10 years a detector like IceCube 
  should observe, for the WB flux, a few such events, 
  whereas no event is expected even for the maximal value of 
  the allowed prompt neutrino flux.

\begin{acknowledgments}
This work was supported by Funda\c{c}\~ao de Amparo
\`a Pesquisa do Estado de S\~ao Paulo (FAPESP), Conselho
Nacional de  Ci\^encia e Tecnologia (CNPq), DOE grant No.  
DE-FG02-95ER40896 and in part by the Wisconsin Alumni 
Research Foundation
\end{acknowledgments}



\newpage 

\vspace{1cm}

\end{document}